\documentclass[a4paper]{jpconf}
\usepackage{graphicx}
\usepackage{wrapfig}
\def\dru{keV$^{-1}$kg$^{-1}$day$^{-1}$}
\begin{document}
\title{Dark matter search project PICO-LON}

\author{K.Fushimi$^1$, H.Ejiri$^{2}$, R.Hazama$^{3}$, H.Ikeda$^4$, K.Imagawa$^5$,
K.Inoue$^4$, G.Kanzaki$^6$, A.Kozlov$^7$, R.Orito$^1$, T.Shima$^2$, Y.Takemoto$^7$,
Y.Teraoka$^4$, S.Umehara$^2$, K.Yasuda$^5$, S.Yoshida$^8$ \\
(PICO-LON Collaboration)}
\address{$^1$ Institiute of Socio, Arts and Sciences, Tokushima University, 1-1 Minamijosanjimacho Tokushima city, Tokushima 770-8502, JAPAN}
\address{$^2$ Research Center for Nuclear Physics, Osaka University, 
10-1 Mihogaoka Ibaraki city, Osaka 567-0042, JAPAN}
\address{$^3$ Graduate School and Faculty of Human Environment, Osaka Sangyo University, 3-1-1 Nakagaito Daito city, Osaka 574-8530, JAPAN }
\address{$^4$ Research Center for Neutrino Science, Tohoku University, 6-3 Aramaki Aza Aoba 
Aoba ward Sendai city, Miyagi 980-8578, JAPAN }
\address{$^5$ I.S.C. Lab., 7-7-20 Saito Asagi Ibaraki City, Osaka 567-0085, JAPAN}
\address{$^6$ Graduate School of Integrated Arts and Sciences, Tokushima University, 1-1 Minamijosanjimacho Tokushima city, Tokushima 770-8502, JAPAN}
\address{$^7$ Kavli Institute for the Physics and Mathematics of the Universe (WPI),
The University of Tokyo Institutes for Advanced Study, The University of 
Tokyo, 5-1-5 Kashiwanoha Kashiwa city, Chiba 277-8583, JAPAN}
\address{$^8$ Department of Physics, Osaka University, 1-1 Machikaneyama
Toyonaka city,  Osaka 560-0043, JAPAN}

\ead{kfushimi@tokushima-u.ac.jp}
\begin{abstract}
The PICO-LON project aims at search for cold dark matter by means of highly radio-pure and 
large volume NaI(Tl) scintillator.
The NaI powder was purified by chemical processing to remove lead isotopes and selecting
a high purity graphite crucible.
The concentrations of radioactive impurities of $^{226}$Ra and $^{228}$Th were effectively reduced to
$58\pm4$ $\mu$Bq/kg and $1.5\pm1.9$ $\mu$Bq/kg, respectively.
It should be remarked that the concentration of $^{210}$Pb, which is crucial for 
the sensitivity to dark matter, was reduced to $24\pm2$ $\mu$Bq/kg.
The total background rate at 10 keV$_{ee}$ was as low as 8 \dru, which was sufficiently low to search
for dark matter. 
Further purification of NaI(Tl) ingot and future prospect of PICO-LON project is discussed.
\end{abstract}

\section{Outline of PICO-LON project}
PICO-LON (Pure Inorganic Crystal Observatory for LOw-background Neutr(al)ino) aims at
search for WIMPs by means of highly radio-pure NaI(Tl) scintillator.
NaI(Tl) scintillator has great advantage to searching for WIMPs because all the nuclei are sensitive to 
both spin-dependent and spin-independent interactions.
The NaI(Tl) scintillator has another advantages to WIMPs search because of  its low background and 
easy to operate under room temperature.

The DAMA/LIBRA group is continuously searching for the signal of WIMPs by highly radio-pure 
and large volume NaI(Tl) crystals \cite{DAMA/LIBRA}.
They developed highly radio-pure NaI(Tl) crystal which contains only a few ppt of U and Th chain
isotope impurities and less than 20 ppb of natural potassium \cite{DAMA_NIM}.
Many other groups are trying to develop highly radio-pure NaI(Tl) crystals to search for WIMPs, 
however, the sensitivity to WIMPs are suffered from a large amount of $^{210}$Pb contamination
\cite{ELEV, ANaIS, DM-ICE, KIMS}.
Recently, the PICO-LON group established the method to reduce $^{210}$Pb in NaI(Tl) crystal.
One of the most serious origin of background was successfully removed and further purification and 
low background test was done.

The final set-up of the PICO-LON detector is planned to consist of 42 modules of large 
volume NaI(Tl) detectors, each with 12.70 cm$\phi\times$12.70 cm. 
The total mass of the detector system is enough to test the annual modulation signal which is 
reported by DAMA/LIBRA \cite{DAMA_TAUP2015}.
The NaI(Tl) crystal is viewed by one photomultiplier tube (PMT) in order to lower the background
events from PMTs.

In the following sections, we will present the recent progresses on the crystal purification and
the result of test measurement of low background measurement.

\section{Development of low background NaI(Tl) scintillator}

The purification of NaI(Tl) ingot is the most important task to develop the high sensitivity detector
to search for WIMPs because
radioactive impurities (RI) in the NaI(Tl) crystal reduces the sensitivity to the WIMPs seriously.
The impurities of RIs in a crystal scintillator should be less than a few tens of $\mu$Bq/kg
in order to use the crystal for dark matter search.
The contamination of $^{210}$Pb is the serious backgrounds because it emits low energy beta rays
($E_{max}=17$ keV and 63.5 keV), the low energy gamma ray and the conversion electron
($E_{\gamma}=46.5$ keV) and L-X rays below 16 keV. 
The $^{210}$Bi, the progeny of $^{210}$Pb, emits high energy beta ray ($E_{max}=1162$ keV) 
which produces bremsstrahlung photons.
All the radiations associated with $^{210}$Pb severely reduce the sensitivity to WIMPs signal.

Although it is quite difficult to reduce the concentration of $^{210}$Pb, 
we have successfully reduced its concentration by chemical process of raw NaI powder.
We tried to remove the Pb ion in the raw powder of NaI by cation exchange resin 
which was optimized to remove the Pb ion.
The raw NaI powder was dissolved in ultra pure water with the concentration of 
300 g/Liter.
The NaI solution was poured into a column in which the cation exchange resin was filled.
The best parameter was searched for and determined to optimize the reduction of 
lead ion by several trials.
The processed solution was dried by rotary vacuum evaporator.
The vacuum of the evaporator was broken by high purity nitrogen gas to avoid the contamination 
by $^{222}$Rn in the air.
As a result, the concentration of $^{210}$Pb became as small as $24\pm2$ $\mu$Bq/kg.

The U-chain ($^{238}$U and $^{226}$Ra) and Th-chain ($^{228}$Th) were effectively reduced by 
purifying the raw material of a graphite crucible.
The graphite was selected based on results of U, Th and K measurements, however, 
we found the purity of the graphite was not sufficiently good because a significant contamination 
of U-chain and Th-chain were observed.
Further purification of graphite was done by baking the graphite under 3000 K\@.
The concentration of $^{226}$Ra and $^{228}$Th were successfully reduced to 
$58\pm4$ $\mu$Bq/kg and $1.5\pm1.9$ $\mu$Bq/kg, respectively.

\section{Low background measurement in Kamioka underground observatory}

The NaI(Tl) ingot was shaved and polished to make 7.62 cm$\phi\times$7.62 cm cylindrical shape.
A quartz light guide with 4 mm in thickness was glued on the top of the cylindrical NaI(Tl) ingot.
All other surfaces of the ingot was covered with 4 mm thick PTFE reflector to guide 
the scintillation photons to the light guide.
The ingot and the light guide were covered with 0.08 cm thick oxygen free high conductive 
copper (OFHC).

\begin{wrapfigure}{r}{0.5\textwidth}
\centering
	\includegraphics[width=0.9\linewidth, bb=0 0 714 836]{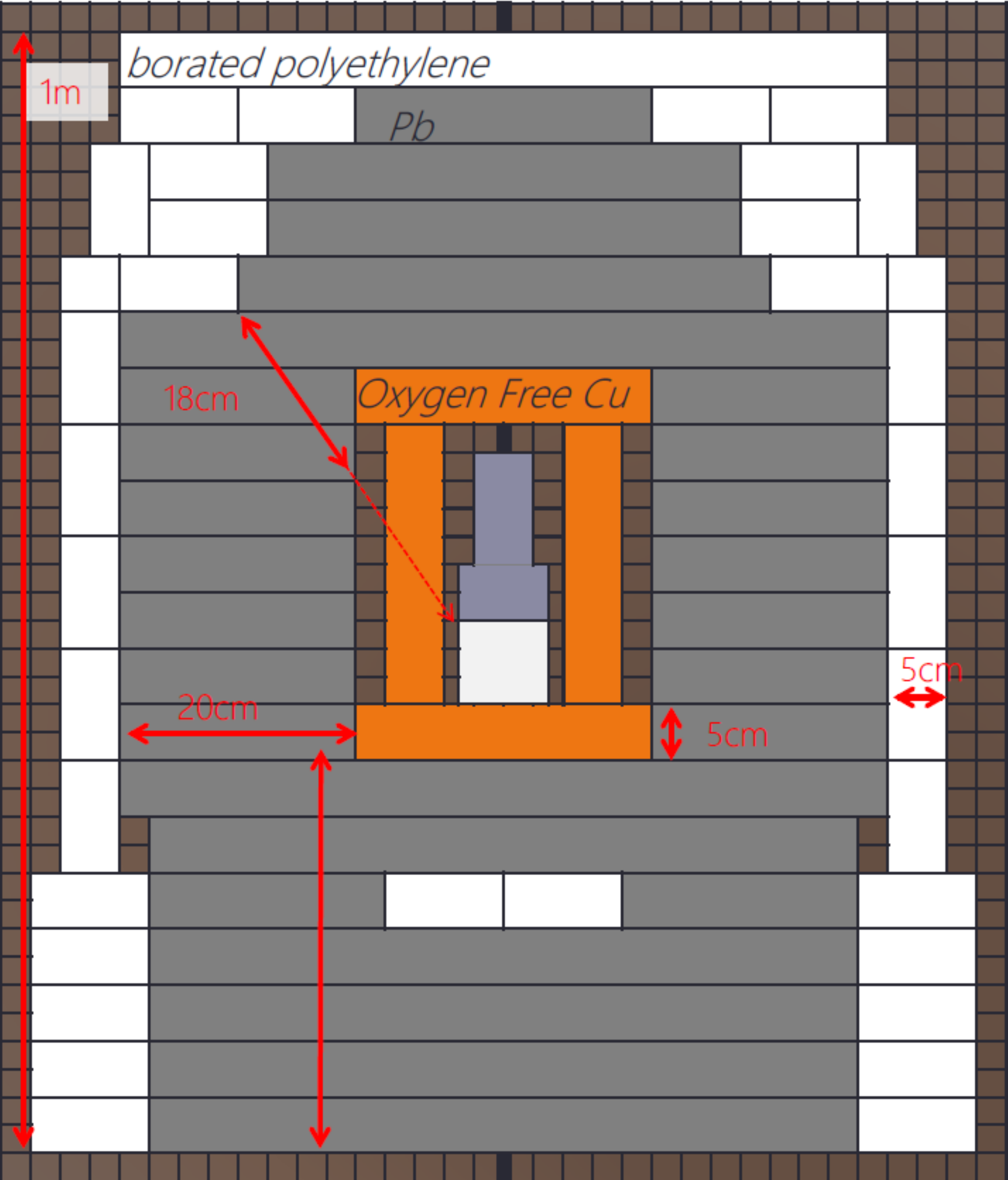}
	\caption{
Geometry of the present measurement in Kamioka underground observatory.
}
\label{fg:Det}
\end{wrapfigure}

The NaI(Tl) detector was covered with 5 cm thick OFHC copper and 20 cm thick old lead passive shield.
No active shield was installed in the present measurement.
The minimum thickness of the lead shield was 18 cm.
Fast neutrons were thermalized and absorbed by 5 cm thick borated polyethylene.
Pure nitrogen gas evaporated from liquid nitrogen was flushed into the inner area of the 
shield to purge radon.
The schematic drawing of the detector system is shown in Figure \ref{fg:Det}.

The low background measurement was started in the summer of 2015  
in Kamioka underground laboratory (36$^{\circ}$25'N, 137$^{\circ}$18'E) located at 2700 m 
water equivalent.
The experiment area was placed in the area of KamLAND experiment.
The air of the experimental room was controlled to keep clean as class 10 by using a HEPA filter.
The flux of the cosmic ray is reduced by a factor of $10^{-5}$ relative 
to the flux in the surface laboratory.

A low background photomultiplier tube (PMT) R11065-20 provided by Hamamatsu Photonics was 
attached on the light guide by optical grease.
The concentrations of U and Th chain in the PMT were less than 10 mBq/module.
The quantum efficiency was as large as 30 \% at the wavelength of 420 nm.

The PMT output pulse was introduced into the fast data acquisition system MoGURA (Module for
General Use Rapid Application)\cite{MoGURA} to digitize the pulse shape.
The trigger for the data acquisition system was produced by timing filter amplifier (TFA)
which integrates 200 nsec.
The fast noise pulses below single photoelectron signals are effectively removed by introducing
TFA and the trigger rate was reduced by about two order of magnitude.

Energy calibration for higher energy range was performed by using $^{133}$Ba 
and $^{40}$K (KCl) sources.
The energy resolution at 1.46 MeV was 6.9 \% in full-width-half-maximum (FWHM).

\begin{wrapfigure}{r}{0.5\textwidth}
\begin{center}
\includegraphics[width=0.89\linewidth, bb=0 0 1463 1040]{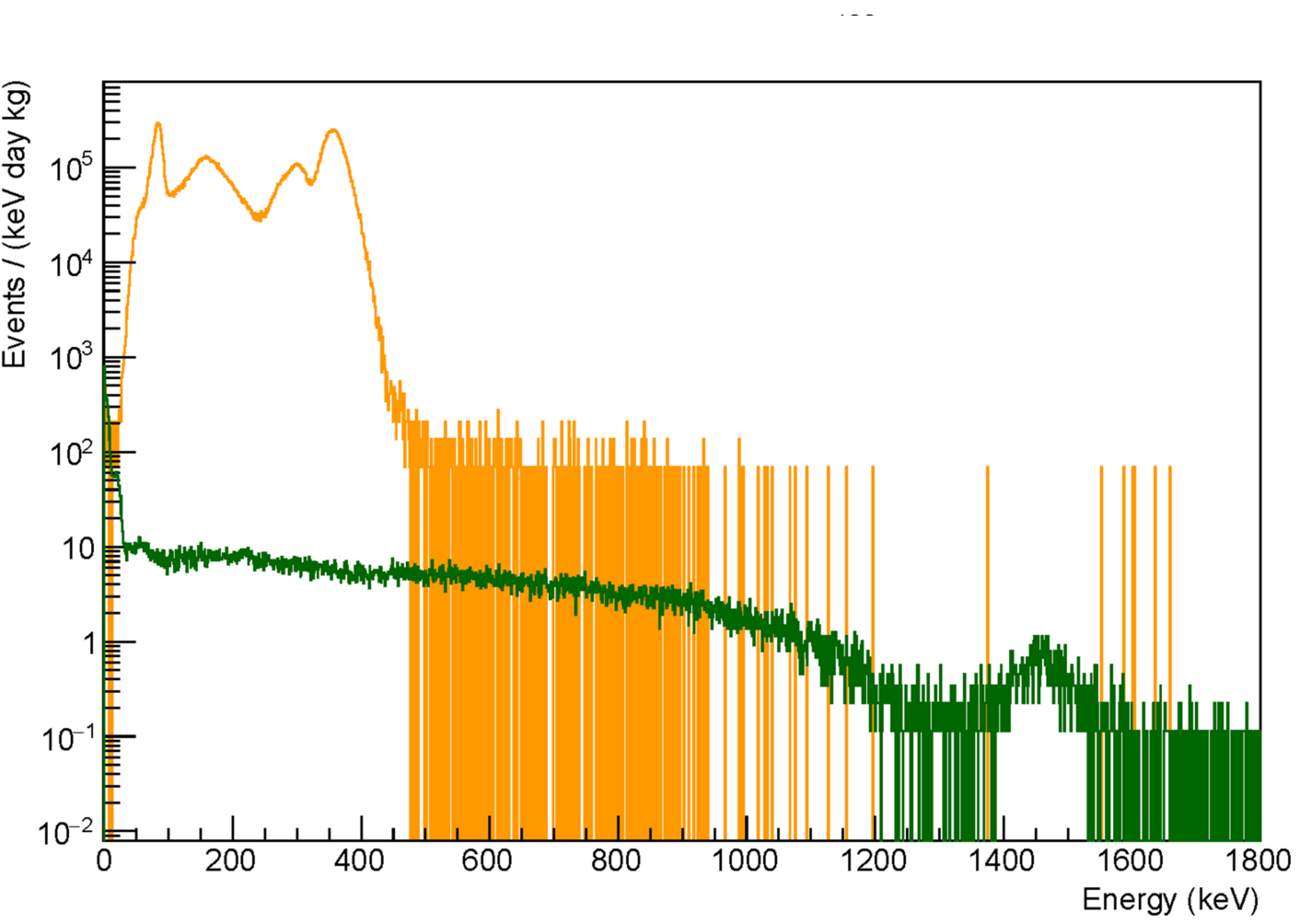}
\end{center}
\caption{\label{Spe}The energy spectra obtained by irradiating $^{133}$Ba (upper orange) 
and background (lower green). }
\end{wrapfigure}
Low background measurement was continued for the live time of 7 days$\times$ 1.2 kg.
The energy spectra of energy calibration and low background measurements are shown in 
Figure \ref{Spe}.
The background energy spectrum was well reproduced by Monte Carlo simulation with the concentration
of the RIs in the surrounding materials.
The present energy threshold was 10 keV$_{ee}$ and the event rate was 
8 \dru at the energy threshold.

\section{Future prospects}
We developed highly radio-pure NaI(Tl) crystal to search for cosmic dark matter.
The RIs of U-chain and Th-chain were sufficiently reduced by purification of 
the raw NaI powder and the graphite crucible.
The significant potassium impurity was observed in the low background measurement.
The Monte Carlo simulation agreed with the assumption that the 2.6 ppm of potassium was contained in 
NaI(Tl) crystal.
The concentration of potassium was too large to use the crystal to the dark matter search.
The chemical process to remove the potassium in NaI raw powder is now in progress.

The background from the surrounding materials is the next important issue.
All the materials which will be used for the detector are selected by measuring the gamma rays from 
the samples.
We started the collaboration with the XMASS group to lower the background from PMTs.
Extensive search for the low background materials will be finished in the beginning of 2016 
and low background PMT will be developed for PICO-LON in 2016.

Full background simulation of 250 kg PICO-LON setup is now ongoing. 
The detail of the detector design is fixing by discussing with Horiba and Hamamatsu Photonics.
The detector design will be optimized to ensure the background rejection by making unti-coincidence 
measurements of background events such as potassium,
1461 keV gamma ray and 3 keV X ray.

\section{Acknowledgment}
The authors thank Professor S.Nakayama for fruitful discussion and encouragement.
The authors also thank Kamioka Mining and Smelting Company for supporting 
activities in the Kamioka mine and  Horiba Ltd. for making the NaI(Tl) detectors.
This work was supported by Grant-in-Aid for Scientific Research (B) number 24340055,
Grant-in-Aid for Scientific Research on Innovative Areas number 26104008.
The work was also supported by Creative Research Project in Institute of Socio, Arts and Sciences,
Tokushima University.
The corresponding author thanks Nogami Fund at RCNP Osaka University 
for the travel support to attend TAUP 2015.

\end{document}